\documentstyle[epsfig]{l-aa}

\def\M10{{\times 10^{10} M_{\odot}\ }}

\begin{document}
\voffset -2.5truecm

\thesaurus{12.12.1;11.09.3;11.17.1}
\title{Evolution of a hot primordial gas
in the presence of an ionizing ultraviolet background:
instabilities, bifurcations, and the formation of Lyman limit systems}

\author{Jan P. M\"ucket \and Ronald E. Kates }
\institute{ Astrophysikalisches Institut Potsdam, \\
An der
Sternwarte 16, D-14482  Potsdam, Germany }
\date{}
\offprints{J. M\"ucket}
\maketitle
\markboth{Instabilities, bifurcations, and Lyman limit systems}{}

\begin{abstract}
The time evolution of thermal and thermo-reactive instabilities of
primordial gas in the presence of ionizing UV radiation is studied.  We
obtain conditions (depending on density, temperature, and UV radiation
intensity) favorable for the formation of a multi-component medium.
Nonlinear effects, especially those attributable to opacity, can play an
important role.  In certain parameter regimes, dramatic, rapid evolution
of ionization states away from ionization equilibrium may occur even if
system control parameters evolve slowly and the system starts in or near
ionization equilibrium.  Mathematically, these rapid changes may best be
understood as manifestations of bifurcations in the solution surface
corresponding to ionization equilibrium.  The astrophysical significance
of these instabilities is that they constitute a nongravitational
mechanism (thermo-reactive instability) for decoupling isolated spatial
regions with moderate initial fluctuations from their surroundings, even
under the influence of heating processes due to ionizing UV radiation.
In particular, isolated, decoupled regions with relatively high neutral
hydrogen density may result, and these could be associated with Lyman
alpha clouds of high column density or with Lyman limit systems.
\end{abstract}

\begin{keywords}
Lyman limit, thermal instability, structure formation
\end{keywords}

\section{Introduction}

An understanding of processes occurring in diffuse gas interacting with a
photoionizing flux is of direct importance in modeling a number of
astrophysical phenomena, such as quasar absorption clouds (Doroshkevich,
M\"ucket, and M\"uller, 1990; Ferrara \& Giallongo, 1996) and the
structure of neutral hydrogen clouds in galactic halos (Ferrara \&
Field, 1994).  The effects of heating due to photoionization in galaxy
formation have been considered by Efstathiou (1992) in the context of
mechanisms for suppression of dwarf galaxy formation and by Navarro and
Steinmetz (1996) in their study of the ``overcooling" problem.  These
authors found that photoionization alone could not provide the heating
mechanism required, i.e., supernova feedback should also be included.
Cosmological simulations of galaxy formation including star formation
and supernova feedback have been carried out by Yepes, Kates, Klypin, \&
Khokhlov (1996; hereafter YKKK).  A reasonable hypothesis is that
photoheating indirectly influences star formation by regulating the
conditions for thermal instability and thus the formation of a
multiphase (cloudy) medium. 

Since many of the observable phenomena involving photoionization require
a prediction of statistical properties
of objects, it is important to integrate the
local dynamics of diffuse gas (i.e., heating, cooling, ionization,
formation of neutral hydrogen) into an overall approach including
hydrodynamics and the evolution of large-scale structure.  As these
processes involve a large dynamical range of scales, some of them will
inevitably occur below the limits of resolution of a numerical
simulation.  Due to the presence of nonlinearities, the effects of
fluctuations below a typical cell size will require special attention,
especially when instabilities are involved.

It has long been appreciated (Field 1965; Defouw 1970; Balbus 1986; Fall
\& Rees; 1985; Iba\~nez \& Parravano 1982) that gas subject to cooling
(and perhaps heating) processes in a cosmological setting may become
thermally unstable, leading to enhanced cooling and the formation of a
multiphase medium including cool clouds (Begelman \& McKee, 1990).  Cool
clouds play an important role in theories of the interstellar medium
(McKee \& Ostriker, 1977) and galaxy dynamics, due to their effects on
the energy budget (enhanced cooling), star formation and
"supernova feedback" (YKKK).

Corbelli \& Ferrara (1995) have demonstrated the existence of so-called
"thermo-reactive" instability modes for gas containing metals in the
presence of ionizing radiation.  Here we will see that instabilities of
thermo-reactive type are also possible in gas of primordial composition.
In particular, our results imply that thermo-reactive instabilities could
play an important role in the formation of at least some population of
observed Lyman limit systems.

The principal goal of this paper is to study the dynamics of fluctuations in a
gas of primordial composition 
which may already have been compressed to
high ambient density (compared to the background density of the universe) and
heated to high temperature as a consequence of large-scale structure formation.
The intention is to characterize instability regimes approximately as a function
of the ambient temperature and density of the gas, with local effects of 
gravitation excluded.  Such a characterization
is expected to be useful for an understanding
of conditions
for star formation in the context of hydrodynamical numerical simulations.  
For this
application, linear instability analysis about ionization equilibrium solutions
does not tell the whole story:  For one thing, even if present, some linear
instability modes are too slow to be of importance on a dynamical timescale.
Moreover, some of the instabilities which occur involve nonlinearity in an
essential way and could not be detected by linear analysis.  (Because opacity is
involved, the strength of the nonlinear effects increases with the size of the
perturbed region; see also Ferrara and Field (1994).)

Apart from its importance for the intended applications, the nonlinear
system studied in this paper is quite interesting from a purely
mathematical point of view:  As seen here [and as previously pointed out
for example by Petitjean, Bergeron, and Puget (1992)], the equations of
ionization equilibrium exhibit multivalued solutions in certain regimes
of parameter space.  We propose that these solutions are best understood
in terms of the theory of {\it bifurcations} (or catastrophes):
Bifurcations are typical in the equilibria of nonlinear systems, and the
mathematical theory (see for example Arnol'd, 1979; Chow \& Hale, 1985)
gives strong hints and indications of the general behavior to be
expected.  Near bifurcations, a system can depart rapidly from
equilibrium even if the control parameters vary slowly.  
(In the present case, the ``slowly-varying" control 
parameters are any two thermodynamic variables, say  
temperature and -- on a still longer time scale -- pressure.)
Hence, even a qualitative
understanding of system evolution near bifurcations requires a time -
dependent treatment of the coupled system of ionization, heating, and
cooling equations.  The importance of simulating the dynamical equations
for evolution of ionization states was previously discussed for example
by Cen (1992), and a dynamical treatment has been incorporated into the
hydrodynamic code used in a number of papers (Cen et al.
1990,1992,1993,1994).  However, for numerical efficiency it would be
useful to know under what circumstances ionization equilibrium is in
fact a good approximation.

The organization of this paper is as follows:  Section \ref{equations}
gives the model system of equations describing the dynamics of the three
species H$_I$, He$_I$, and He$_{II}$ for a gas of primordial
composition.  Section \ref{catas} shows that the solution manifold of
the ionization equilibrium equations can exhibit the properties of a
mathematical bifurcation surface (Golubitsky \& Schaeffer, 1985).  Of
special interest are the solution trajectories of constant pressure, for
which one sees that evolution through equilibrium solutions is not
always possible.  Section \ref{evoldens} considers selected solutions of
the full time-dependent equations exhibiting some typical behavior.
Section \ref{appl} characterizes the regimes of temperature and
density in which instabilities
relevant to numerical simulations (i.e., those with sufficiently short
timescales) are most likely to occur.  Section \ref{concl} summarizes
and describes in particular some potentially interesting implications for
Lyman limit systems.

\section{Equations describing thermal and ionization evolution}
\label{equations}

Consider first the time-evolution of a sufficiently small
region such that the density
$\rho(t)$ and gas temperature $T(t)$ may be treated
as approximately uniform.  (This approximation
will be discussed below in connection with
the optical depth $\tau_\nu$.)
Self-gravity is neglected.
The dynamics in this region are assumed to depend on Hubble
expansion, cooling processes, and heating due to a UV background.
The UV background is
assumed to be described by an externally generated,
explicitly time-dependent, spatially
uniform flux $J(t)$.

We restrict our consideration to a region containing
hot gas with primordial abundances,
i.e. the total helium number density $n_{He}$ is 10 \% of the total
hydrogen number density $n_H$.  The medium is characterized in addition
by the number densities $n_{HI}$,$n_{HII}$,$n_{HeI}$,$n_{HeII}$ and
$n_{HeIII}$ of its ionization states HI, HII, HeI, HeII, and HeIII,
respectively.  One also has $n_H=n_{HI}+n_{HII}$ and
$n_{He}=n_{HeI}+n_{HeII}+ n_{HeIII}$ and the equations of charge
conservation.

The evolution of these number densities is determined by the equations

\begin{equation}
{dn_{HI}\over dt} =-\xi_{HI} n_{HI} -
\xi_{e,HI} n_{HI} n_e +n_{HII} n_e \alpha(HII) \label{e2HI}
\end{equation}

\begin{eqnarray}
{dn_{HeI}\over dt} &=&-\xi_{HeI} n_{HeI} -
\xi_{e,HeI} n_{HeI} n_e \nonumber\\
                   & & +n_{HeII} n_e \alpha(HeII) \label{e2HeI}
\end{eqnarray}

\begin{eqnarray}
{dn_{HeII}\over dt} &=&-\xi_{HeII} n_{HeII} -
\xi_{e,HeII} n_{HeII} n_e\nonumber\\
                    & &+n_{HeIII} n_e \alpha(HeIII) -{dn_{HeI}\over dt}
\label{e2HeII}\end{eqnarray}
In the above equations, $\xi_i$ and $\xi_{e,i}$ [with $i$ ranging over
(HI,HeI, HeII)] denote the rates of photo- and collisional ionization,
respectively, $n_e$ and $n_i$ are the electronic and ionic number
densities, and $\alpha(i)$ denotes the recombination rate from species
$i$ to the next lower ionization state.  The photo-ionization rate
$\xi_i$ for the $i$-th species is written in the form

\begin{equation}
\xi_i(x) = \int_{\nu i}^\infty 4\pi{J_\nu(x) \sigma_i(\nu)\over h\nu} d\nu
\label{e3}
\end{equation}
where the $\nu_i$ are the threshold frequencies of each species,
$J_\nu$ is
the intensity of the diffuse background UV flux in units of
ergs  s$^{-1}$ cm$^{-2}$ Hz$^{-1}$
sr$^{-1}$, and $\sigma_i(\nu)$ is the photoionization cross
section, which is approximated using the expressions given by
Kramers (1923).

In addition to the primary ionizations described by Eq.  (\ref{e3}),
secondary ionizations could contribute to the number density evolution
(Shull, 1979; Shull \& van Steenberg, 1985; Ferrara \& Field 1994).
However, there is observational evidence for a steepening of the UV
spectrum at the HeII edge (Jakobsen et al., 1994; Ferrara \& Gialongo,
1996).  Hence, the energy of most secondaries will be low enough that a
large fraction, though not all, of the energy will be converted to heat.

The diffuse flux is diminished by absorption due to surrounding gas
layers (self-shielding).  The optical depth at a distance $x$ from the
surface is given by
\begin{equation}
\tau_\nu = \int_0^x dx' \sum_i n_i(x') \sigma_\nu(i)
\label{int_tau}
\end{equation}
where in the summation $i$
ranges over all three species for \mbox{$\nu_{\rm HeII} \le \nu $},
over HeI and HI for \mbox{$\nu_{\rm HeI} \le \nu < \nu_{\rm HeII}$} and
over HI for \mbox{$\nu_{\rm HI} \le \nu < \nu_{\rm HeI}$}.

We model the UV flux in Eq. (\ref{e3})
as a power law spectrum with
spectral index $\alpha$; it may thus
be expressed in the form

\begin{equation}
J_\nu(x) = J_0(\nu/\nu_{LL})^{-\alpha} \exp (-\tau_\nu(x))
\label{fluxlaw} \end{equation}
where $J_0$  is the
intensity at the Lyman limit; we take $\alpha=1$ in the
calculations that follow.

For later use, it will be convenient to introduce a shielding depth
for HI defined by
\begin{equation}
\tau_{HI} \equiv -\log(\xi_{HI}/({4\over 3}\pi J_0 G^B_{HI})),
\label{irate1}
\end{equation}
where $G^B_{HI}$ is the
parameter calculated by Black (1981) for the optically thin case.
(Shielding depths for HeI and HeII can be defined similarly.)

A knowledge of the above quantities allows us to obtain the heating
rates by photoionization

\begin{equation}
\Gamma_i(x) = \int_{\nu i}^\infty 4\pi {J_\nu(x)h(\nu - \nu_i)\over h\nu}
 \sigma_i(\nu) d\nu
\label{hrate}
\end{equation}
and to construct the total heating rate
\begin{equation}
\Gamma = \sum_i \Gamma_in_i .
\end{equation}
The cooling rate $\Lambda$ is calculated according to Black (1981),
and Compton cooling is included.

Now, the flux $J_\nu$ appearing in the photoionization rates (\ref{e3})
and the heating rates (\ref{hrate}) is a monotonically decreasing
function of $x$, which differs from an exponential because of the 
$x$-dependence of $\tau_\nu$, which itself depends on the $n_i(x)$.  As we
will see below, even in a region with moderate gradients and slowly
evolving thermodynamic state variables (pressure and temperature) the
number densities $n_i$ of the ionization states can have large gradients
and can also change rapidly.  Nonetheless, because the $x$ - dependence
of (\ref{e3}) and (\ref{hrate}) is the same, one may always define
weighted means $n_i$ such that the integral in Eq.  (\ref{int_tau}) is
replaced by
\begin{equation}
\tau_\nu = x \sum_i n_i \sigma_\nu(i)
\label{constni}
\end{equation}
where the summation is as above and Eqs. (\ref{e3}) and (\ref{hrate})
remain accurate as written. Of course, these
weighted means $n_i$ still depend on $x$, and they are not
strictly equal to the average number densities.

Here we identify the true average number densities
with these weighted values and use them in the
evolution equations (\ref{e2HI}), (\ref{e2HeI}), (\ref{e2HeII}).
For the optically thin case, this is an excellent approximation,
whereas in the optically thick case there are
some quantitative (but not qualitative)
inaccuracies as discussed in the Conclusions.

\section{Bifurcation structure in the
manifold of solutions for ionization equilibrium}
\label{catas}

One can gain some insight into the time-dependent problem by studying
the system representing ionization equilibrium, which is obtained by
setting all time derivatives to zero in Eqs.  (\ref{e2HI}),
(\ref{e2HeI}) and (\ref{e2HeII}).  (For the purposes of this discussion,
the environment is considered static, i.e., constant flux intensity
$J_0$, no cosmological expansion.)  For a range of number densities
$n_H$ and temperatures $T$, the complete set of possible solutions for
$n_{HI}, n_{HeI}$ and $n_{HeII}$ was identified and found numerically.
Each solution corresponds to some net heating (or cooling) rate
$\Gamma-\Lambda \propto \dot{T} \equiv dT/dt$, and the locus of
equilibrium solutions defines a two-dimensional surface embedded in a
three-dimensional space with coordinates ($n_H$, $T$, $\Gamma-\Lambda$).

\begin{figure}
\begin{minipage}[b]{.4\linewidth}
\epsfig{file=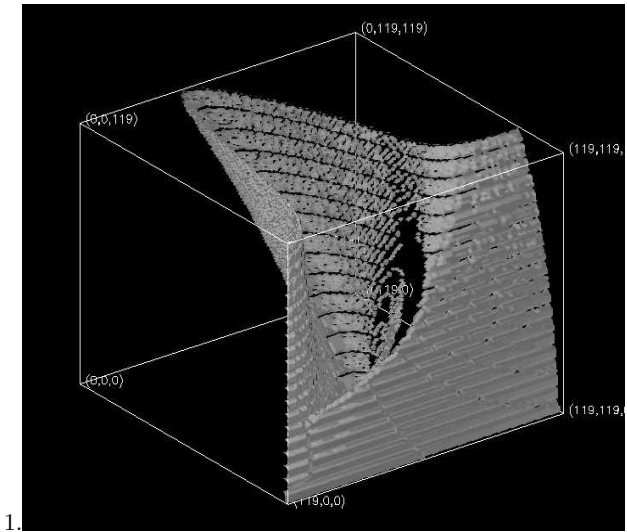,scale=1., height=7cm,width=8cm}
\end{minipage}
\vglue0.5cm
\begin{minipage}[b]{.4\linewidth}
\epsfig{file=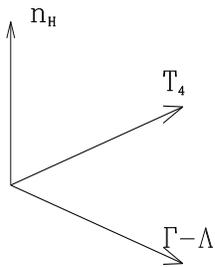,scale=1., height=4cm,width=4cm}
\end{minipage}
\caption[]{3-dimensional surface plot of $[ \Gamma-\Lambda]$
 as a function of $n_H$ and $T$ in
 the range $10^{-4}cm^{-3}<n_H<2 \times 10^{-3}cm^{-3}$ and $1<T_4<5$.
 The surface is folded and contains at least one bifurcation
 (axes as indicated in the lower panel).}
\label {kata3d}
\end{figure}

Fig.  \ref{kata3d} shows a plot of this surface in the range
$10^{-4}cm^{-3}<n_H<2 \times 10^{-3}cm^{-3}$ and $1<T_4<5$, where $T_4$
is the temperature in units of $10^4$K.  The flux intensity was $J_0 =$
$J_{21}$ $10^{-21} $ ergs s$^{-1}$ cm$^{-2}$ Hz$^{-1}$ sr$^{-1}$, with $J_{21}
\approx 0.05$.  This flux density provides a typical example of a
surface containing a classical catastrophe or bifurcation fold which
appears to terminate in one or more cusps of the generic type described
by Whitney (1955):  For temperatures near $1.3 < T_4< 2.1$ (the precise
range depending on the density within the regime $5 \times
10^{-4}<n_H<10^{-3}$) there are three solutions for a given $n_H$, $T$.

The existence of a bifurcation has profound consequences.  First, as a
general property, any system is essentially nonlinear near a
bifurcation, and therefore there is no hope of understanding the
resulting instabilities from a linearized analysis.  Secondly,
as control parameters of a (time-dependent) system initially in a state
of equilibrium are slowly varied, departures from equilibrium are path
dependent:  if the path passes through a region with no folds, the
system may evolve quasistatically through a family of nearly ionization
equilibrium states.  However, near a bifurcation even a slow
change in control parameters forces the system to undergo such rapid
evolution that time-dependent solutions would be expected to depart
substantially from (ionization) equilibrium.

For a qualitative illustration of what to expect from time dependent
simulations and for further understanding of the bifurcations, it is
useful to study the behavior of $\dot{T}=dT/dt$ (or equivalently
$\Gamma - \Lambda$) along curves of constant
pressure $P$ but changing $T$.
The isobaric trajectories define a one-parameter family
of hyperbolae through the two-dimensional ($n_H$, $T$) parameter space.
The time-evolution of $n_H$ and $T$ for this restricted system
corresponds to moving along one of these hyperbolae either in the
direction of decreasing $T$ (cooling) or increasing $T$ (heating) at the
rate given by $\dot{T}$.

\begin{figure}
\epsfig{file=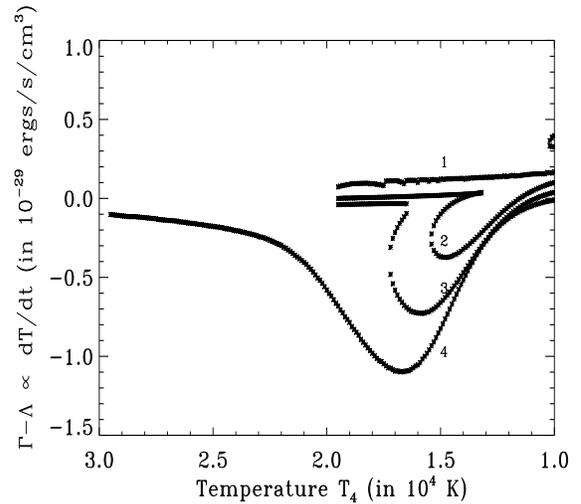,scale=1., height=7cm,width=8cm}
\caption[]{Energy loss ($\Gamma - \Lambda $) for different UV fluxes
at constant pressure  $n_H T = 0.95\cdot 10^2$ K cm$^{-3}$.
}
\label {fig2}
\end{figure}

Fig. \ref{fig2} shows a family of plots of $\Gamma-\Lambda$ vs.  $T$
along
hyperbolae defined by $P$=const for varying flux coefficients (from top
(1) to bottom (4)) $J_{21}$=[0.05, 0.01, 0.005, 0.001].  As the flux
decreases, the curves have deeper minima.  Beginning on the left, as
long as $\Gamma-\Lambda<0$, an isobaric system would move along the
curve to the right (decreasing temperature).  For the low-flux curve (4), the
temperature drops monotonically until thermal equilibrium is reached
somewhere below $10^3$~K.  The only other noteworthy feature in this
temperature range is a higher cooling rate near $T_4=1.7$.

\begin{figure}
\epsfig{file=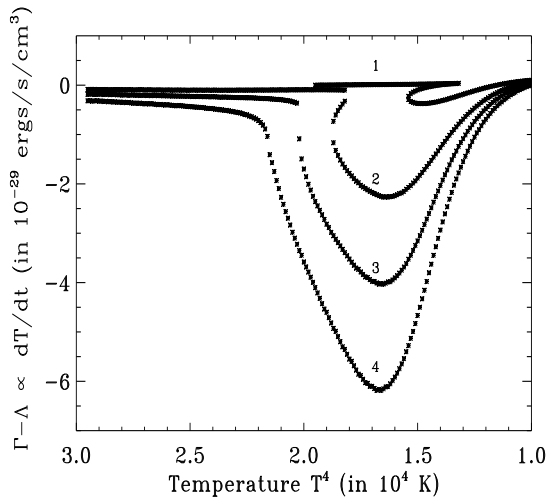,scale=1., height=7cm,width=8cm}
\caption[]{Energy loss ($\Gamma - \Lambda)$ for various
constant pressures and constant UV flux $J_{21}$ = 0.01.
Curve (1): $P$ corresponds to $T_4=0.95$ at $n_H=10^{-3}$;
Curve (2): $P=$ 5 times higher than curve (1)
Curve (3): $P=$ 10 times higher than curve (1)
Curve (4): $P=$ 50 times higher than curve (1)
}
\label {fig3}
\end{figure}

The next higher flux ($J_{21}$=0.005) curve (3) is triple valued between
$T_4=1.7$ and the cusp near $T_4=1.6$.  The three solutions correspond
to three different ionization states solving the equilibrium equations.
With respect to the full time-dependent equations, the two solutions with
the highest and lowest values of $n_{HI}$ are stable (attractive), while
the middle solution is unstable (repulsive).  Moreover, this regime is
likely to be reached by a time-dependent system, because $dT/dt$ is
negative up to the cusp.

The next higher flux curve (2) also has a triple-valued regime, but the
curve crosses the $T$ axis (thermal equilibrium) near $T=1.4$ and thus
is not {\it forced} to the cusp.  Nevertheless, three solutions of the
ionization equilibrium equations are available in the range where the
system is still cooling.  If some perturbation (such as a small
variation in the flux or external pressure) occurs, it is possible for
the system to jump to the lower stable solution, cool, and evolve to
lower temperatures.

The upper curve in Fig. \ref{fig2} includes a triple-valued solution within the
heating regime (see upper right).  A system initially at $T_4<1$ would
be heated until reaching the cusp near $T_4=1$ and could then jump to
the lower branch with a lower heating rate.

The curves in Fig. \ref{fig3} are similar to those of Fig. \ref{fig2},
except
that the
pressure is varied for constant flux.  In the lower curve (4), the
solutions are single-valued and do not enter the heating regime.  The
next higher curve (3) has a very steep gradient near $T_4=2.0$, where it
passes rather close by the bifurcation.  The next curve (2) is analogous
to curve (3) in Fig. \ref{fig2}. Following the upper curve (1) from left to
right, a system at strictly constant pressure would stabilize near
$T_4=2$ before the multi-valued region is reached.  However, slight
departures from constant pressure could move the system to the
multi-valued region and permit a rapid change to the other stable
branch.

The behavior of the curves in Figs. 2 and 3 is quite typical of systems
with a bifurcation.  In particular, they illustrate the difficulties
likely to arise in a numerical simulation.

For all of the multi-valued curves of Figs. 2 and 3, the solution
with the largest cooling rate is also associated with a drastically
higher fraction of neutral hydrogen.  In particular, at a given pressure
it is possible for the system to have both an optically thin and an
optically thick stable state.  This possibility gives a hint that there
could be structures associated with the existence of qualitatively
different phases in neighboring regions with initially similar (but not
identical) control parameters.

\section{Evolution and instability of perturbed regions in an
ambient hot primordial gas} \label{evoldens}

In the previous section, we saw that an isobaric system may cool in such
a way that it is forced out of ionization equilibrium.  Moreover, we
wish to study the evolution of a system evolving so quickly that the
control parameters $T$ and $n_H$ need not change simply quasistatically,
because the timescales associated with cooling and/or expansion may be
short compared to the relaxation time to ionization equilibrium.  Both
of these problems require solution of the time-dependent equations
(\ref{e2HI} - \ref{e2HeII}).

\begin{figure}
\epsfig{file=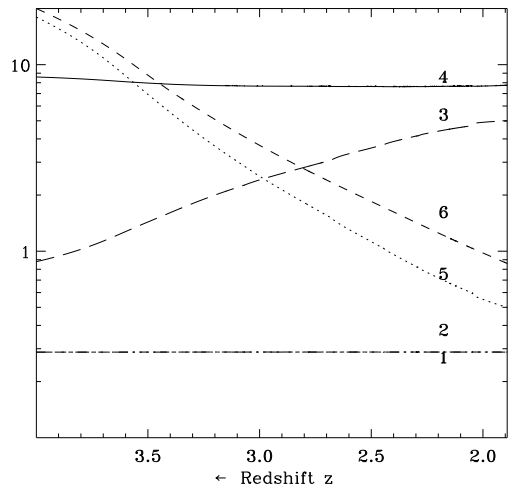,scale=1., height=7cm,width=8cm}
\caption[]{Evolution of characteristic parameters 
as a function of the redshift $z$ for initial
conditions $z_i=4$, $T_h=20$, $n_h=0.75 \times 10^{-3}$~cm$^{-3}$,
$J_{21}(z_i)=0.5$. (The flux $J_{21}(z)$ evolves according to 
Fig. \ref{logflux}.)
Starting from the lower left of the
figure:  Two bottom lines: shielding depth $\tau_{HI}$ (definition
see Eq. (\ref{irate1}) for the hot
(1; dot-dot-dot-dash) and cold (2; dot-dash) components.
 Long dashed curve (3): ratio (hot-to-cold) of Jeans lengths.
Solid curve (4): $-\log_{10}(n_{HI}/{\rm cm}^{-3})$ for the cold component.
Dotted curve (5): $T_c$ in units
of $T_4$.  Short dashed line (6): $T_h$ in units of
$T_4$. }
\label {fig4}
\end{figure}

\begin{figure}
\epsfig{file=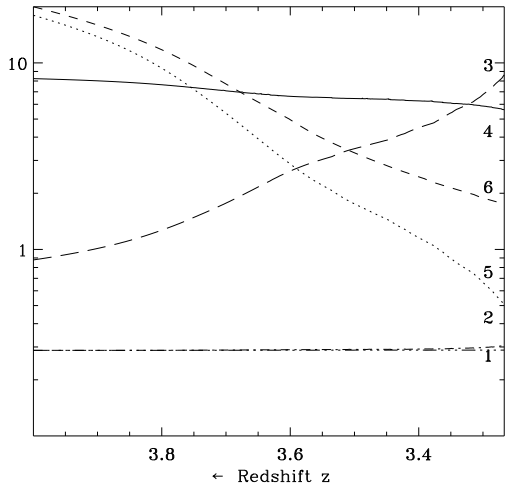,scale=1., height=7cm,width=8cm}
\caption[]{Evolution as in Fig. \ref{fig4},
except that the density is
increased by a factor of 2.
Labeling as in
Fig. \ref{fig4}.}
\label {fig5}
\end{figure}

\begin{figure}
\epsfig{file=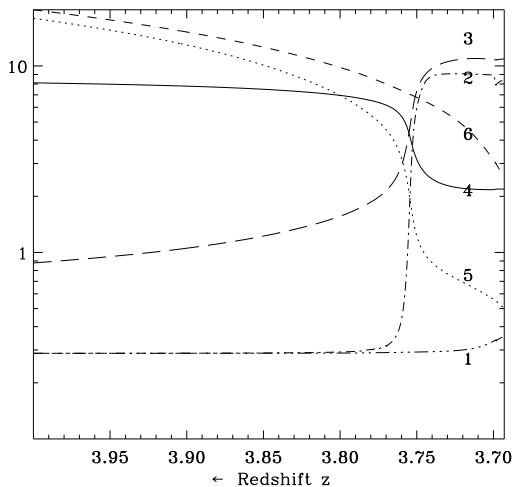,scale=1., height=7cm,width=8cm}
\caption[]{Evolution as in Fig. \ref{fig5},
except that the initial flux is
reduced by a factor of 10 to $J_{21}=0.05$. Labeling as in
Fig. \ref{fig4}.}
\label {fig6}
\end{figure}

We are interested in the evolution of structures in the intergalactic
medium and particularly in the formation of cold clouds.  These effects
may occur in a variety of environments, including both gas which is
still comoving with the Hubble flow and gas in structures that have
decoupled from the expansion.  In this paper, we derive results for the
former case (comoving); however, as discussed below, the effect of
decoupling the gas from the expansion would usually tend to enhance the effects
studied here.

In general, the equations of cooling and ionization are coupled to the
gas dynamical equations, as well as to gravitation.  However, the
qualitative character of the instabilities can be studied without the
full generality of gas dynamics.  The "one-cell" dynamics considered
here could be generalized to include advection within the context of a
numerical simulation that does include gas dynamics and gravitation (see
for example YKKK).

We consider a gas with
primordial composition consisting of two components:
\begin{itemize}
\item {an ambient (``hot") component
at temperature $T_h(t)$, with uniform number density $n_h$
 (not to be confused with $n_H$, the
hydrogen density).  The gas is assumed as explained above to be
comoving with the cosmic expansion.}

\item
{an initial perturbation (the ``cold" component) consisting of a slightly
compressed region with number density $n_c$ (assumed spatially constant in the
perturbed region) evolving in pressure equilibrium with the hot component and
therefore at a lower temperature $T_c=T_h n_h/n_c$.  For concreteness, we
consider the cold perturbation region to be
a flat slab of thickness $50$ kpc. However, subject to the
approximation of Eq. (\ref{constni}), the results should
apply to a range of geometries.}
\end{itemize}
We take a ratio of baryons to dark matter
of $\Omega_B=0.1$ in both components and ambient density
contrast $\delta$ initially with respect
to an assumed background number density of
$2 \times 10^{-7}$~cm$^{-3}(1+z_i)^3$.  This implies $n_h$ is initially
$ (2 \times 10^{-7}$ cm$^{-3}) (1+z_i)^3(1+\delta)$.

In many treatments of thermal instability such as Fall \& Rees (1985),
Field (1965), one is primarily interested in whether or not
a perturbation can decouple.  Here, decoupling is only of interest
if it occurs on a time scale short compared to
$t_{\rm grav} \propto t_{\rm Hubble}/\sqrt(1+\delta)$.

Suppose that the gas is {\it initially} in ionization equilibrium.
Assuming an initial perturbation of $10\%$ at some initial
redshift $z_i$, we study instabilities by following the evolution during
the time $t_{\rm grav}$.  During the evolution we monitor
parameters characterizing possible decoupling of the system
including
\begin{itemize}
\item {the ratio $\rho_c/\rho_h$}
\item {the temperatures $T_c$ and $T_h$  }
\item {the fractions $n_{HI}$ of neutral hydrogen }
\item {the optical depths $\tau_c$ and $\tau_h$}
\item {the "cooling enhancement" ratio $E_c/E_h$, where
for each component $E$ is the energy lost (gained)
per unit volume}
\item {the multiplicity of the solutions of ionization equilibrium,
if applicable}
\item {the Jeans lengths for each component}
\end{itemize}

As disussed by Cen (1992) in a more general context,
an important technical problem is the presence of
several varying time scales, including $t_{\rm grav}$, the
cooling times $t_{\rm cool}$, and the various time scales
associated with the time-dependence of the three species
$n_{HI}$, $n_{HeI}$, and $n_{HeII}$ in Eqs. (\ref{e2HI} - \ref{e2HeII}).
The complete time dependent equations are solved at all times using
adaptive timestep control taking into account the {\it shortest} relevant timescale.
However, it sometimes occurs that
the time-dependent solution is close to an  (ionization) equilibrium solution.
It is then computationally efficient to use a different
algorithm which finds an appropriate equilibrium solution
and tests that this solution is indeed close to
the time-dependent solution sought.
In this case, the timestep is comparable to the
dynamical time scale $t_{\rm dyn}$, the minimum of
$t_{\rm grav}$ and $t_{\rm cool}$.

Fig.  \ref{fig4} shows the evolution with redshift of characteristic parameters
for initial conditions $z_i=4$, $T_h=20$, $n_h=0.75 \times 10^{-3}$~cm$^{-3}$,
$J_{21}=0.5$.  This density corresponds to 30 times the background density.  For
these parameters, no thermo-reactive instability occurs during a time $>>
t_{grav}$ ( $t_{grav} \approx 10^8$~years), and the cold component does not
decouple from the environment:  The shielding depths $\tau_{HI}$ for the hot and
cold components coincide in the figure.  Both
components remain optically thin throughout.  The curve (3) giving the ratio of
the Jeans length of the hot component to that of the cold component attains
a value of about 5 during the course of the simulation and has
flattened out considerably by the end of the simulation at $z=1.9$. Neither
component has become Jeans unstable within the time
scale considered.  

Fig.  \ref{fig5} is for the same parameters as in Fig. \ref{fig4}, except that
the density is increased by a factor of 2.  This figure illustrates 
how an instability may indeed be present (i.e., according to
linear perturbation theory) whose timescale for decoupling however
is longer than the dynamical timescale.  
Clearly, some degree of decoupling has taken place: the ratio of Jeans
lengths (hot to cold) is 10 and still growing, and  $T_h$ is about 4 times
 $T_c$. As before, the 
shielding depths $\tau_{HI}$ for the hot and cold components 
are practically indistinguishable. 

In contrast to the above two figures, 
Fig.  \ref{fig6} shows a clear case of decoupling.
The parameters are as in Fig. \ref{fig5}, except that
the initial flux has been reduced by a factor of 10 to $J_{21}=0.05$.  
Even in this case
the optical depths of the cold and hot component remain very close
until about $z=3.77$.  However, subsequently, the optical depth of the
hot component barely changes, remaining optically thin as in Figs.
\ref{fig4} and \ref{fig5}, 
whereas the cold perturbation becomes optically thick
($n_{HI}$ increases by more than four orders of magnitude) and at the
same time rapidly cools and thus drastically decreases its Jeans length,
until it becomes gravitationally unstable.  Note that the rapid increase
in neutral hydrogen density at $z=3.75$ is not a numerical artifact, but
represent rapid evolution as in Figs. \ref{fig2} and \ref{fig3}
where the solutions of the
static ionization equilibrium equations are multi - valued.  This
evolution provides an example of nonlinear thermo-reactive instability
and results in a strong decoupling of the cold perturbation from its
environment.

\section{Investigation of instability regimes in the parameter
space of density and temperature}\label{appl}

As mentioned above, one important goal of this paper is to determine what
regimes of ambient density and temperature are favorable for eventual star
formation in a region containing primordial gas.  The arguments given here
apply to ``star formation" as it occurs in the context of the 
McKee - Ostriker (1977) theory of the
interstellar medium.  Now, the formation and evolution of the interstellar
medium are certainly complicated processes involving various modes of
instability on a range of spatial and time scales.  However, thermal or
thermo-reactive instabilities are expected to be a necessary prerequisite for
such a multiphase medium to arise in the first place.  Hence, a characterization
of those regimes of ambient gas temperature and density likely to produce
thermo-reactive or thermal instabilities will give information on environments 
favorable for star formation.  
(Since instability depends on UV flux density as well as
density and temperature, the flux will obviously play an important role in
regulating star formation.)

To this end, we have simulated the evolution of ``cold" and warm
components in pressure equilibrium as in Section \ref{evoldens} under a
wide range of initial conditions in number density and temperature, and
also for a variety of realistic flux and redshift combinations.  All
simulations began with an initial gas density perturbation of 10~\% and
were continued until either the colder component cooled to below 5000~K
or the local Hubble time $t_{\rm grav}$ defined above was exceeded.
(Some perturbations which would be classified as ``unstable" according
to linear perturbation analysis would nonetheless fail to cool within
this time.)  The dimension of the perturbed region was taken to be
$50/(1+z)$~kpc.

\begin{figure}
\epsfig{file=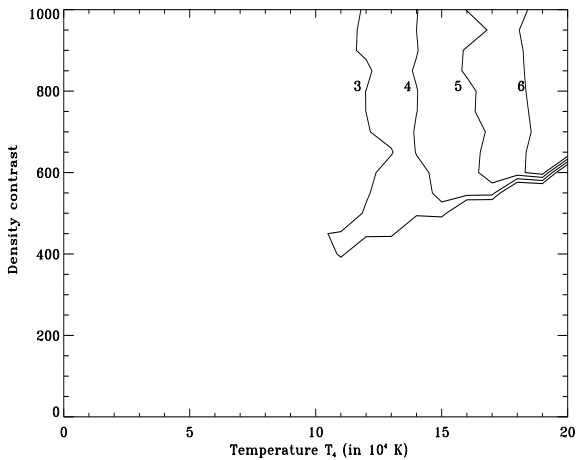,scale=1., height=6.5cm,width=8cm}
\caption[]{ Contour plot of the ratio ${\rho_c/\rho_h}$ evaluated at
the end of the simulation; initial redshift $z_i=0.1$ and initial flux
$J_{21}=0.01$.   Horizontal axis: $T_4$ for the
hot component. Vertical axis:
initial overdensity of  hot component with respect
to the background density. Contour labels: ${\rho_c/\rho_h}$}
\label {c2}
\end{figure}

We now present detailed results of several simulations on a
finely-spaced, two-parameter grid (temperature-number density).  We
first consider the case of initial redshift z=0.1 and flux
$J_{21}=0.01$:  Fig.  \ref{c2} is a contour plot of the ratio
${\rho_c/\rho_h}$ evaluated at the end of the simulation.  The
horizontal axis is the initial temperature of the hot component in units
of $10^4$~K, and the vertical axis represents the initial overdensity of
the hot component (with respect to the mean {\it gas} density).  The
visible contours represent regimes in which the cold component actually
did succeed in cooling.  In the present discussion, we regard a final
ratio ${\rho_c/\rho_h}>3$ as an indicator of decoupling (instability).
Note that the density ratio could have been higher than its final value
during the course of the simulation (see Fig.  \ref{fig5} near
$z=3.75$).  One may observe from the figure that the regime of
instability as characterized by this indicator consists of the region
roughly defined by the inequalities $T_4 \ge 10$ (a
vertical boundary) and
\begin{equation}
n_{\rm gas,init}>(400+25 T_4) n_{\rm gas,0},
\label{aympt1}
\end{equation}
where $n_{\rm gas,0} \approx 2 \cdot 10^{-7}$~cm$^{-3}$
is the average gas density today.

\begin{figure}
\epsfig{file=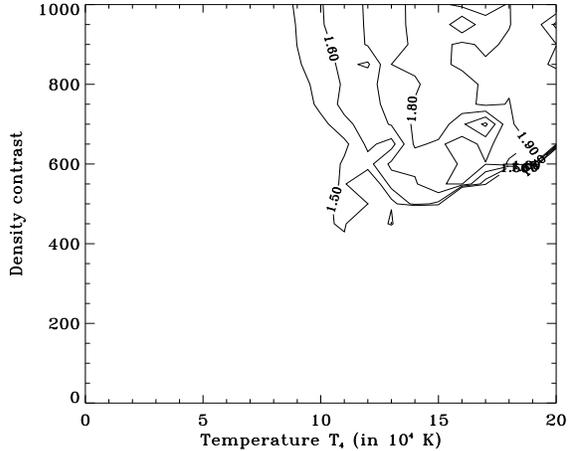,scale=1., height=6.5cm,width=8cm}
\caption[]{Contour plot of enhancement (see text for definition);
axes as in Fig. \ref{c2}}
\label {c3}
\end{figure}

Fig.  \ref{c3} is a contour plot of the ``enhancement" ratio of the cold
to the hot component (ratio of total integrated heat lost per unit mass
during the simulation for the respective components) for the same simulations as
in Fig. \ref{c2}.  Again, although
the final enhancement ratio may be lower than the maximum during the
course of the simulation, it still may be regarded as an indicator of
instability and decoupling.  The regime of decoupling as defined
by this indicator (enhancement greater
than 1.5) in Fig.  \ref{c3} coincides quite well with that defined by
the final density ratio (Fig.  \ref{c2}).  Again, the visible contours
are plotted only if the perturbed gas actually cooled.

\begin{figure}
\epsfig{file= 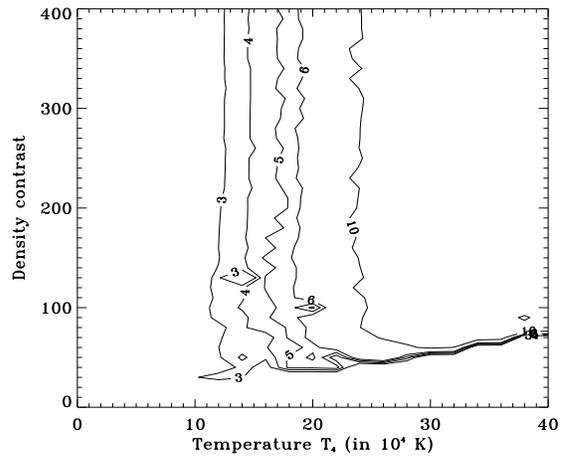,scale=1., height=6.5cm,width=8cm}
\caption[]{Contour plot of the final value of the
ratio $\rho_c/\rho_h$ as in Figs. \ref{c2}, but
for initial redshift z=3 and initial flux $J_{21}=0.1$}
\label {cz3}
\end{figure}

Since the UV flux changes with redshift, the instability regimes and
thus the conditions for development of a multicomponent medium will
evolve.  Hence, one cannot necessarily estimate instability conditions in the
distant past simply by extrapolating the low-redshift results to take
into account the higher background density.  We next report the results
of higher-redshift simulations performed as before on a fine-parameter grid,
but for the following redshift-flux combinations:
\begin{quote}
$z_i=3, J_{21}=0.1$ (see Fig.  \ref{cz3}) and \\
$z_i=4$, $J_{21}=0.5$ and $J_{21}=1.0$.
\end{quote}
These fluxes are all within the constraints provided by observations and
large-scale structure simulations (M\"ucket et al., 1996; Bechtold,
1994; Williger et al., 1994; Bajtlik, Duncan, \& Ostriker, 1988; Lu,
Wolfe, \& Turnshek, 1991; Haardt \& Madau, 1995).  It is quite
remarkable that for these higher initial redshifts, the left
(temperature) boundaries of the instability regimes are nearly unchanged
from the low-redshift case and also hardly differ for the two flux
values ($J_{21}$) considered:  the temperature limit is still roughly
characterized by $T_4 > 10$.

At redshift $z_i=3$ for an initial flux $J_{21}=0.1$, one can see from
Fig.  \ref{cz3} that the bottom (density) boundary of the instability
regime can be approximated by
\begin{equation}
n_{\rm gas,init}(z_i=3)>(25 + 1.8T_4)*n_{\rm gas,0} \cdot (1+z_i)^3.
\end{equation}
At the initial redshift $z_i=4$
for the lower flux value (0.5), the regime of instability is roughly
characterized by the condition
\begin{equation}
n_{\rm gas,init}(z_i=4)>(14+1.1T_4)*n_{\rm gas,0} \cdot (1+z_i)^3.
\end{equation}
For the higher value
($J_{21}=1.0$), the lower density bound is shifted upward by an
additional factor of about 1.3.

\begin{figure}
\epsfig{file=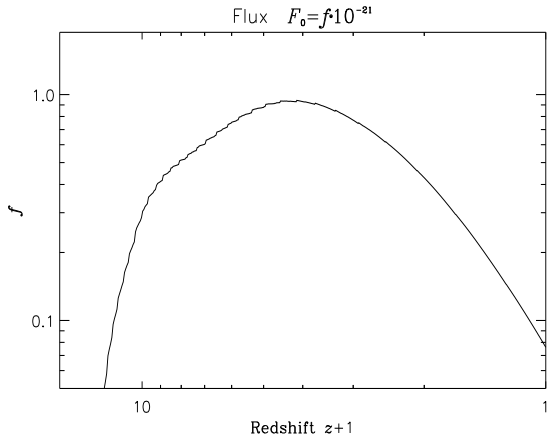,scale=1., height=6cm,width=8cm}
\caption[]{Log-log plot of redshift dependence of the UV flux obtained
from N-body simulations; here the flux refers to
$f = 4\pi J_{21}$}
\label {logflux}
\end{figure}

In order to investigate the general trend of evolution with redshift for
the asymptotes described above, we have analyzed the instability
behavior on a coarser grid with respect to ($T_i, n_{H,i}$ ) for a
sequence of redshifts $z_i = 0.1, 1, 2, 3, 4, 5$ and flux values
$J_{21,i}$.  As a model for the redshift dependence of the UV flux, we
have used the results provided by simulations described in M\"ucket et
al.  (1996a) which reproduce the observations mentioned above quite well
(see Fig.  \ref{logflux}).  The results are given in Table \ref{table1}.
For each initial redshift $z_i$, the asymptotes (as defined previously)
are given approximately by an expression of the form
\begin{eqnarray}
T_4 &=& T_0 = const,\nonumber\\
n_{\rm gas,init}(z_i)&=&(n_0 + m_0 T_4)*n_{\rm gas,0} \cdot (1+z_i)^3
\label{as1}
\end{eqnarray}

\begin{table}[htb]
\centerline{
\begin{tabular}{|ccccc|} \hline
$z_i$ & $J_{21}$&$T_0$&$n_0$&$m_0$ \\
\hline
0.1&    0.008&  12&     400&    5\\
1.&     0.034&  10&     89&     3\\
2.&     0.061&  8&      44&     1.75\\
3.&     0.074&  10&     25&     1.1\\
4.&     0.073&  10&     17&     0.65\\
5.&     0.06&   10&     11&     0.5\\
\hline
\end{tabular}
}
\caption{The parameters $T_0, n_0, m_0$ for the
asymptotes defining the regions of instability at given initial
redshift $z_i$ and flux $J_{21,i}$ in the $n-T$ plane.}
        \label{table1}
\end{table}

From these simulations,
one may deduce that the parameter $n_0$ is most strongly
determined by requiring that the flux be almost
entirely shielded: Since we consider a planar
slice of fixed comoving size $x_0=50$ kpc, this 
requirement would imply that the critical value $n_0$ 
is determined by the relation 
$\tau = x_0 \sigma n_{HI} (1+z_f)^2$, 
where $z_f$ is the redshift
for which $n_{HI} \approx n_0 n_{\rm gas,0}  = n_H$, i.e., 
when hydrogen becomes neutral.  The results of Table \ref{table1}
roughly fit this relation for a critical
optical depth on the order of $\tau \approx 10 - 15$ (taking into
account typical differences between $z_i$ and $z_f$).
The values $n_0 (1+z_i)^2$ in 
Table \ref{table1} vary within 20\% around the
value 400 (the scatter is probably attributable to the coarse
computational grid of steps in $T$ and $n_H$ in the numerical
calculations).  There is also a weak (roughly logarithmic) dependence on
the flux, as one would expect.  

At higher temperatures ($T_4 > 10$) the terms describing
collisional ionization become important.  The stage at which the flux
is shielded ($\tau \approx 10$) and cooling becomes strong enough
is therefore shifted to higher densities $n_H$.  Although the functional
dependence is certainly more complicated, the slope $m_0 (1+z)^3$ can be
roughly estimated as being proportional to the flux $J_{21}$ up to
redshift $z=4$.

\begin{figure}
\epsfig{file=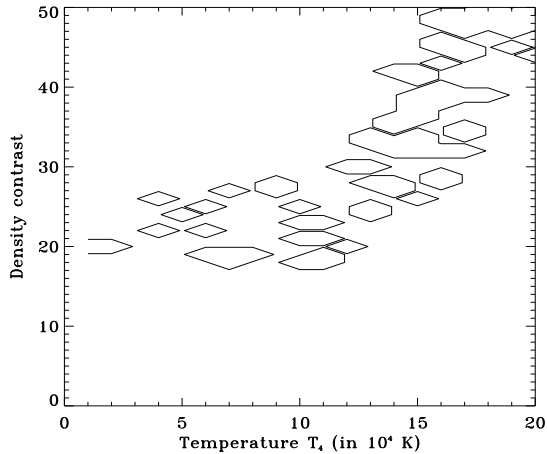,scale=1., height=6.5cm,width=8cm}
\caption[]{Simulations as in Figs. \ref{c2} and \ref{c3}:
regimes in which hot gas gained net energy,
but cold perturbation lost energy.}
\label {c1}
\end{figure}

We now return to the simulations discussed above for $z_i=0.1$ and
$J_{21}=0.01$.
In contrast to the decoupling criteria illustrated in Figs.
\ref{c2},\ref{c3}, and \ref{cz3}, in which both cold and hot
components always lost energy, in the low to moderate density contrast
regime it is possible for decoupling to occur in which the hot gas
component actually {\it gains} energy, while the cold perturbation loses
energy (cools).
This effect is seen in the contours of Fig.
\ref{c1},
which form a rough band in the figure.  Above this band, there is a net
energy loss in both components; below this band, there is a net energy
gain in both components.  The somewhat ``speckled" behavior within the
band is due to the fact that energy may be lost and gained within the
course of a simulation.  The existence of this band could be of great
interest in a scenario in which the spatial dependence of shielding is
treated more precisely than here.  In this case, there could be a {\it
persistent} core of opaque gas (high neutral hydrogen) surrounded by
hotter, optically thin gas which is heated (See Ferrara \& Field, 1994).
Above the band, one would expect the core to increase gradually in size
(as in a cooling flow), whereas below the band a perturbation might be
expected to evaporate, especially if heat conduction is properly
included.

In order to get an idea of the length scales of the structures involved,
we apply the
results of Cowie \& McKee (1977) to the parameters considered here and
obtain an expression for typical values of the critical evaporation
timescale $t_{\rm evap}$, which involves the length scale.  A critical
evaporation length of 100 pc may then be estimated from the condition
$t_{\rm evap} > t_{\rm cool}$.  A larger limit is obtained from the
condition $t_{\rm evap} > t_{\rm grav}$, which leads to a critical size
of 1 kpc.  This length scale also follows from a different line of
argument given by Ostriker \& Gnedin (1996).

\begin{figure}
\epsfig{file=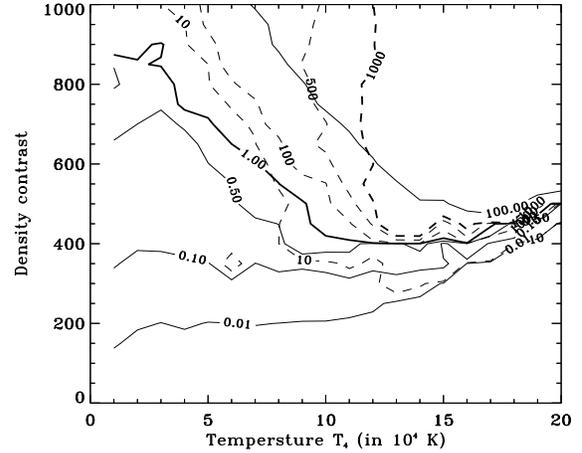,scale=1., height=6.5cm,width=8cm}
\caption[]{Simulation parameters $z$, $J_{21}$ and
axes as in Figs. \ref{c2} - \ref{c1}:
Solid contours: final optical depth of  cold perturbation.
Dashed contours: neutral hydrogen
number density ratio $R_HI \equiv n_{HI,c}/n_{HI,h}$.   }
\label {c_gesamt}
\end{figure}

Fig.  \ref{c_gesamt} is a contour plot showing the final optical depth
of the cold perturbation (solid line) and the neutral hydrogen number
density ratio of cold to hot components $R_{HI}$ as a function of
initial ambient density and temperature (for initial redshift z=0.1 and
flux $J_{21}=0.01$ as before).  The regions (upper right in Fig.
\ref{c_gesamt} with both opacity $\tau > 1$ and $R_{HI}>>1$ coincide
with the instability regions identified in Fig.  \ref{c2} on the basis
of a moderately elevated density contrast.  They represent the
thermo-reactive instability as seen earlier
for one example of evolution in Fig.  \ref{fig6}.  The instability
manifests itself as a dramatic rise in $\tau$ and $R_{HI}>>1$.  The
interpretation will be discussed below.

\section{Conclusions}
\label{concl}
We have obtained regimes of thermal and thermo-reactive instability for
a range of initial redshifts and fluxes (see Sect. \ref{appl} and Table
\ref{table1}).
In each case, the boundaries of
the instability regime as defined by independent 
criteria (enhancement, density contrast, etc.) were mutually
consistent
and only moderately dependent on the instability thresholds defined.  
As seen in Section \ref{appl}, it is
remarkable that even in the presence of flux there is a lower
temperature bound to the instability regime that is
almost independent of the density as long as the
lower bound is exceeded.  
The minimum (initial) number density for instability to occur
corresponds to requiring that the flux be almost
entirely shielded.  

In order to estimate the quality of the approximation and averaging procedure
in Eq. (\ref{constni}), we
have performed a spatially dependent 
analysis for selected critical cases.  This analysis 
shows that the qualitative picture does
not change and the error is far below one order of magnitude.  Moderate
quantitative changes would be expected chiefly with respect to the
characteristic sizes obtained.  The advantages of the simplifying
assumption made at the end of Section \ref{equations} are that they
allow one to obtain insight into the qualitative behavior of optically
thick systems and that the analysis, which involves an exhaustive search
of parameter space, can be provided for reasonable computing times.

The above characterization of instability regimes
favorable for star formation is of particular use 
in the context of hydrodynamical numerical simulations 
of large-scale structure with galaxy formation (YKKK; Steinmetz, 1996).
In such codes, smoothed estimates of the gas
density and temperature are produced at every timestep.  
Consider for the present discussion a medium with
primordial abundances; suppose that some estimate of the ultraviolet
flux is available (see e.g. M\"ucket et al.  (1996)). 
Now, the evolution of thermal and thermo-reactive instabilities involves the
ionization states.  These in turn depend in principle not only on the {\it
current} local values of the temperature, density, and ultraviolet radiation
flux, but also on the past history.  However, considering the computational
expense of explicitly simulating the time evolution of all relevant ionization
species in every cell, it is useful to study what can be inferred from a
knowledge of the current pressure, density, and ionizing flux alone.  The
present paper shows that it should generally be possible to estimate the
instability regimes for any desired redshift/flux combinations by making use of
look-up tables compiled ``off-line" at each redshift as a function of
temperature and pressure.  The trends found here at least roughly characterize
conditions favorable for formation of a multiphase medium as a function of
redshift and the ambient temperature and density.

Such a characterization of instability regimes would be an
extremely useful tool in studying the influence of photoheating on star
formation, and hence on galaxy evolution.  The suppression of thermal
instability with increasing UV flux also illustrates how an antibiasing
mechanism might arise if one were to take into account explicitly the
emission and transport of UV radiation by massive galaxies and quasars
(Efstathiou, 1992; Haardt \& Madau, 1995; Ferrara \& Giallongo, 1996).

Referring to Figs.  \ref{fig6} and \ref{c_gesamt}, we identify the
thermo-reactive instability seen in the evolution of the cold
perturbation as a possible mechanism for at least some Lyman limit
systems:  During the course of the evolution, the optical depth of the
perturbation increases dramatically according to our model.  For a
region of size 50 kpc, column densities ranging from 10$^{17}$ to about
10$^{20}$ cm$^{-2}$ or neutral hydrogen densities of up to $5\cdot
10^{-3}$ cm$^{-3}$ would result.  (Note that the full hydrogen density
$n_h$ also increases with respect to its initial value.)

Realistically, the evolution of the thermo-reactive instability leading
to Lyman limit systems will need to be modeled more precisely taking
properly the spatial variation of the optical depth and the onset of
Jeans instability into account.  One would expect that the
thermo-reactive instability would have a tendency to propagate out from
an initial core.  Once the instability has affected a region exceeding
the Jeans length, as occurs in Fig.  \ref{fig6}, gravitational
instability would set in.  

The conditions would thus be favorable for
protogalaxy formation leading to star formation and hence
some degree of enrichment of the medium with 
heavy elements at a later stage. Thus, one would expect 
Lyman limit systems that had previously
formed from primordial gas according to our scenario  
to contain heavy elements when observed now.  
Indeed, evidence for heavy elements in Lyman limit systems
has been observed 
(Petitjean \& Bergeron, 1990; Petitjean et al., 1994).

In Fig.  \ref{c_gesamt} one also finds regimes for which the cold
component is decoupled ($R_{HI} >> 1$) but still optically thin.  These
regions could be of interest in more detailed models of the formation of
Ly-$\alpha$ clouds.

Since the main goals of this paper were to characterize the energy
budget of primordial gas in terms of parameters (density, temperature)
available from numerical simulations, the question of bifurcations was
not addressed here in full mathematical detail.  However, in view of the
potential importance of bifurcations for a qualitative understanding of
observations (such as Lyman limit systems), the general mathematical
structure of ionization problems in a diffuse gas is worthy
of further study in its own right.

Acknowledgements:
We wish to thank Professor Andrea Ferrara for numerous helpful and
clarifying comments.
We are especially grateful to Gustavo Yepes for productive
discussions. REK was supported by a Fellowship
(Ka1181/1-1) from the DFG (Germany).

\def\reference{\bibitem[\protect \citename{}]{ciiaa}~}

\end{document}